\documentclass[10pt,a4paper]{article}
\usepackage{graphicx}
\usepackage{amssymb}

\begin{document}

\title{Doppler Tomograms from Hydrodynamical Models of Dwarf Nova Disks}
\author{A. J. Gawryszczak \and M. R\'o\.zyczka }

\maketitle

\abstract{We present three-dimensional models of accretion disks in U~Gem - like systems and 
calculate their Doppler tomograms. The tomograms are based on two different assumptions 
concerning the origin of line emission from the disk. The assumption of lines originating due to 
irradiation of the surface layer of the disk by the central source leads to a better agreement 
with observations. We argue that fully three-dimensional modelling is necessary to properly 
interpret the observed tomograms.}

\section{Introduction}
With their typical dimensions of less than $10^{-4}$ arcseconds, Dwarf Nova (DN) disks are far too 
small to be resolved directly. However, an indirect observational insight into their structure 
became possible already in mid-eighties, when a powerful technique of Doppler tomography was 
introduced (for a recent review see Marsh 2001). Nowadays tomographic observations are a 
standard, but the interpretation of Doppler tomograms in terms of theoretical models of DN disks 
is often problematic. This is particularly well visible in the case of spiral waves, which have 
been suggested as one of the agents responsible for the angular momentum transfer through the 
disk (for a recent review see Boffin 2001). The excellent discussion of the subject can be found 
in a recent paper by Smak (2001), and there is no need to repeat it here. 

The main problem associated with the interpretation of Doppler tomograms is related to the nature 
of the data derived from the theory. Model calculations yield spiral patterns in the distribution 
of disk surface density, while patterns in observational tomograms are related to the distribution
of the emissivity in specific spectral lines. While comparing these two sets of data one usually 
makes an implicit assumption that emissivity is proportional to surface density, which certainly 
is an oversimplification.

A more sophisticated approach was presented by Steeghs and Stehle (1999), who based their Doppler 
tomograms on emission line profiles calculated from 2D disk models. For the origin of the lines 
they adopted a purely thermal model with local Planckian source function. Such model was 
criticized by Smak (2001), who pointed out that it requires factor of 100 overabundance of helium 
in order to reproduce the observed intensities, while the relative brightness of the features it 
produces in the tomograms is incompatible with observations. 

Smak himself proposed to base the theoretical tomograms on the distribution of velocity 
divergence. He argued that at a given radial distance from the disk center the compression 
regions with $\nabla\vec{v}<0$ would be distinguished by a higher-than-average disk thickness 
(because their density and temperature would be higher). As a result, the surface layer of the 
disk would be better exposed to the irradiating flux from the white dwarf and the boundary layer, 
and a local enhancement in line emission would be observed. Based on the three-body model of gas 
flow in a close binary he identified regions of maximum compression in the orbital plane and 
showed that they well reproduced shape, location and relative intensities of the arch-like 
structures observed in Doppler tomograms of DN disks. 

Motivated by his paper, we obtained two- and three-dimensional hydrodynamical models of DN disks 
and calculated their tomograms. The details of the modelling procedure are given in 
Section~\ref{Methods}. The results are presented in Section~\ref{results} and discussed in 
Section~\ref{disc}.

\section{Numerical methods, input physics  and initial conditions}
\label{Methods}

All models presented here were obtained with the help of the ZEUS-3D code (Clarke \& Norman 1994, 
Clarke 1996). The original code was modified to include conservative angular momentum transport 
(Kley 1998). Spherical coordinates $(r,\theta,\phi)$ centered on the primary and corotating with 
the system were used. The grid was extending from $0$ to $2\pi$ in $\phi$, from $0$ to $0.2\pi$ 
in $\theta$, and from $r_\mathrm{in}=0.1a$ to $r_\mathrm{out}=0.5a$ in $r$, with $a$ standing for 
the orbital separation. Grid spacing was uniform in $(\theta,\phi)$ and logarithmic in $r$, 
resulting in zones of identical shape. After a few experiments we decided to limit the resolution 
to $100\times20\times64$ zones in $r,\ \theta,\ \phi$ directions, respectively (test runs, with 
resolution increased by a factor of 2 in $r,\ \theta,\ \phi$ consecutively, did not introduce
any significant changes into the models). 

A standard periodic boundary condition was imposed at $\phi=2\pi$, and symmetry with respect to 
the orbital plane was assumed, implying a reflecting boundary condition at $\theta=0$. A free 
outflow was allowed for at $r_\mathrm{in}$ and $r_\mathrm{out}$. In the four innermost radial 
zones the radial velocity was reduced by 5\% at every time-step, preventing the reflection of 
waves from the inner boundary of the grid. To check whether this damping procedure did not 
influence the structure of the disk or the shape of the spiral pattern, we calculated model B2 
with $r_\mathrm{in}$ moved to $0.045a$ (see Table~\ref{ModParam}). It was found that shifting the 
inner grid boundary toward the white dwarf did not introduce any significant changes in the model.

The simulations did not include explicit viscosity (the von Neumann \& Richtmyer and scalar linear 
artificial viscosities originally implemented in ZEUS were only used with coefficients 
$C_1=0.5$ and $C_2=2.0$, where $C_1$ is responsible for the magnitude of the artificial viscous 
pressure, and $C_2$ is a shock-spreading parameter, see the definitions in Stone and Norman (1992)). 
The energy equation was not solved; a polytropic equation of 
state ($p=\kappa \rho^\gamma$) with $\gamma=5/3$ was employed instead. In a system of units in 
which gravitational constant, orbital separation, and primary's mass are all equal to 1, the 
value of $\kappa$ was set to 6500. To mimic U~Gem - like systems, all models had the same mass 
ratio, $q=0.5$. The stream flowing from the secondary through the $\mathrm{L}_1$ point was not 
included.

\begin{table}[h]
  \caption[]{List of models}
  \label{ModParam}
  $$ 
  \begin{array}{lllll}
    \hline
    \noalign{\smallskip}
    \mathrm{Model} & \mathrm{type} & n_\mathrm{radial} & r_\mathrm{in}\ & \mathrm{remarks} \\
    \noalign{\smallskip}
    \hline
    \noalign{\smallskip}
    \mathrm{A} & \mathrm{2D} & 100 & 0.1a  & \\
    \mathrm{B1} & \mathrm{3D} & 100 & 0.1a  & \\
    \mathrm{B2} & \mathrm{3D} & 150 & 0.045a & \mathrm{outer\ 100\ zones\ placed\ as\ in\ B1}\\
    \noalign{\smallskip}
    \hline
  \end{array}
  $$ 
\end{table}

Every simulation consisted of three phases (relaxation, switch-on, and proper). The mass ratio was
set to 0 in the relaxation phase and to 0.5 in the proper phase, while in the switch-on phase it
was linearly increasing in time. At the beginning of each simulation ($t=0$) the grid was 
initialized with an exponential density distribution
\begin{equation}
  \label{rhodistr}
  \rho(r,\theta)=\max\left(\rho_0 \mathrm{e}^{-\alpha r^2\sin^2\theta},\rho_\mathrm{min}\right),
  \label{eq:init_density}
\end{equation}
where 
\begin{equation}
  \alpha=\frac{G M_1}{2 r^{3}} \frac{1}{c_\mathrm{s,0}^2},  
\end{equation}
and
\begin{equation}
  c_\mathrm{s,0}^2=\left. \frac{\partial p}{\partial \rho} \right\vert_{\theta=0}= 
        \gamma \kappa \rho_0^{\gamma-1} .
\end{equation}

In our system of units the value of the midplane density, $\rho_0$, was set to $10^{-8}$, 
corresponding to $\sim2\cdot10^{-8}\,\mathrm{g\,cm^{-3}}$ in U~Gem (with U~Gem parameters taken 
from Groot 2001). The azimuthal velocity of the disk was given a purely Keplerian pattern, and 
the remaining two velocity components were set to $0$. Since the content of the grid was not in 
hydrostatic equilibrium, we allowed it to relax for $\sim3.9$ orbital periods ($P_\mathrm{orb}$). 
Throughout the relaxation phase the model was strictly axisymmetric, so that it was possible to 
speed the computations up by reducing the number of angular grid points to 2. 

Because of extremely steep vertical gradients of density at the surface of the disk it was 
necessary to introduce a density limit. Every time step the grid was scanned for cells with 
$\rho<\rho_\mathrm{min}$, and whenever such a cell was found $\rho$ was reset to 
$\rho_\mathrm{min}$. We wanted $\rho_\mathrm{min}$ to be small enough to minimize side-effects 
caused by the newly-added matter falling onto the disk, and, simultaneously, large enough to 
avoid excessive computational slowdowns due to formation of strong shocks in the rarefied medium 
above the disk. After a few experiments $\rho_\mathrm{min}$ was set to $10^{-13}$ for all models. 
At the end of the relaxation phase the midplane density of the disk increased, reaching up to 
$4\cdot10^{-8}$ (corresponding to $\sim8\cdot10^{-8}\,\mathrm{g\,cm^{-3}}$ in U~Gem). 

At the beginning of the switch-on phase the relaxed model was mapped onto the standard grid, and 
the secondary's gravity was ``switched on''. The final value of $q$ was achieved at 
$t=5.8\,P_\mathrm{orb}$. At the end of the switch on phase the total mass contained in the grid, 
and scaled to U~Gem, $M_\mathrm{disk}$, was equal $\simeq10^{24}\,\mathrm{g}$). The proper phase 
with $q=0.5$ lasted for another $\sim3.9\,P_\mathrm{orb}$ so that the simulation was ended at 
$t\simeq9.7\,P_\mathrm{orb}$. By that time shape and location of the disk edge stabilized, and a 
stationary spiral pattern developed in the disk.

\section{Results}
\label{results}

The tidal forces affect the relaxed disk in two ways. First, some material is stripped from its 
outer edge and driven out of the grid through the outer grid boundary. Second, angular momentum 
is removed from the remaining material and transferred into the orbital momentum of the binary, 
causing the disk to shrink. In the three-body approximation, the radius of the disk cannot be 
larger than the radius of the largest non-intersecting orbit, $r^\mathrm{max}_\mathrm{nis}$. 
U~Gem - like systems with $q=0.5$ have $r^\mathrm{max}_\mathrm{nis}\simeq0.3$, and in fact at the 
end of the simulation the disk barely extends beyond $r\simeq 0.3$. The rest of the gas 
originally located at $0.3<r<0.5$ now resides in a ring-like density enhancement between 
$r\simeq0.15$ and $r\simeq0.3$. 

\begin{figure*}[t]
  \centering 
  \resizebox{\hsize}{!}{
    \includegraphics{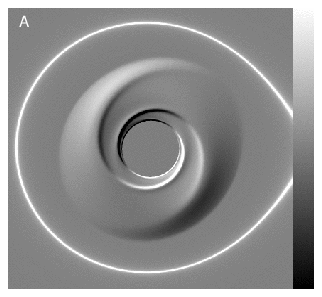}
    \includegraphics{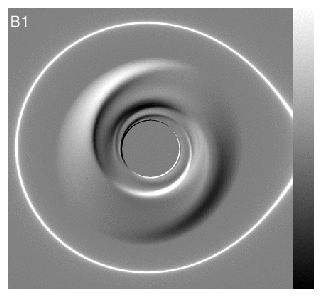}
  }
  \caption{Left panel: distribution of disk density, $\rho$, in model~A. Right panel: distribution
    of disk surface density, $\Sigma = \int\rho(z)dz$, in model~B1. The white loop marks the 
    Roche lobe.}
  \label{densmaps}
\end{figure*}

The ring is markedly elliptical, but, when averaged over the azimuthal angle, it shows a 
well-defined density maximum at $r\simeq0.19$, i.e. slightly beyond the circularization radius 
($r_\mathrm{circ}=0.16$ for $q=0.5$). The maximum density is  factor of $\sim 2.5$ higher than the 
midplane density of the inner disk ($r<0.15$). The ratio $h/r$, where $h$ is the half-thickness 
of the disk, varies from $\sim0.1$ in the inner disk to $\sim0.2$ in the ring. The overall 
structure of the final model is reminiscent of the one expected at the early phase of the 
outburst, when the outer radius of the disk just begins to increase. However, because of the 
polytropic equation of state we employ, the interior of the disk is unrealistically hot (for a 
disk composed of pure hydrogen $T$ grows from $\sim10^4\,\mathrm{K}$ at the surface of the disk to 
$\sim 6.7\cdot10^5\,\mathrm{K}$ at the density maximum). We find that both 2-D and 3-D 
calculations produce disks of nearly the same shape and extent (Fig.~\ref{densmaps}). Below we 
shall argue, however, that fully three-dimensional models are needed to properly interpret the 
Doppler tomograms.

\begin{figure*}[t]
  \centering 
  \resizebox{\hsize}{!}{
    \includegraphics{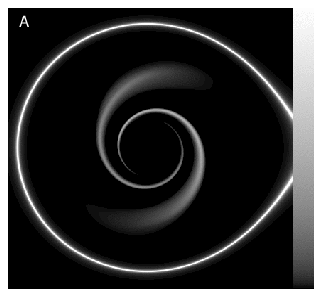}
    \includegraphics{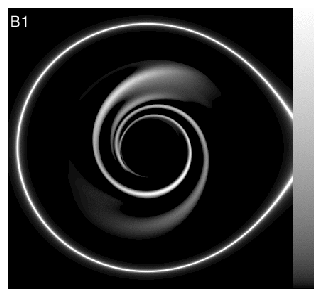}
  }
  \caption{Left panel: distribution of the compressional power per unit volume, 
    $-min(p\nabla\vec{v},0)$, in model~A. Right panel: distribution of the compressional power 
    per unit surface, $-\int min(p\nabla\vec{v},0) \mathrm{d}z$, in model~B1.} 
  \label{miscmaps}
\end{figure*}

Both 2-D and 3-D models develop spiral shocks shown in Figs.~\ref{densmaps} and~\ref{miscmaps}. At 
a first glance, location and inclination of the shocks do not significantly depend on the number 
of dimensions of the model. Significant differences become visible when compression regions 
($\nabla\vec{v}<0$) are compared: while model~A has a clear two-armed pattern, three arms are 
present in models B1 and B2. We speculate that the third arm may originate due to tidal forcing 
of the disk matter in the direction perpendicular to the orbital plane (an effect entirely absent 
in 2D). Thus, to the long dispute on whether spiral shocks can exist in three dimensions, or 
rather their existence is limited to the two-dimensional world, we add a vote in favor of the 
first possibility. In 3D the shocks are definitely there, but their pattern is different than in 
2D. Obviously, the validity of this conclusion is limited to hot polytropic disks with 
$0.1\leq h/r \leq0.2$. 

Following the approach of Smak (2001), we obtained Doppler tomograms of the compressional power 
due to tidal forces, $p\mathrm{d}V$.The distribution of brightness on the $(v_x,v_y)$ plane was 
calculated from the integral
\begin{equation}
  L_{p\mathrm{d}V}(v_x,v_y) = -\int_V\min(p\nabla\vec{u},0) {\cal{B}}(u_x, v_x, \delta_{v_x}) 
  {\cal{B}}(u_y, v_y, \delta_{v_y}) \mathrm{d}V,
  \label{eq:PdV_tomogram}
\end{equation}
where ($u_x, u_y$) are velocity components of the volume element $\mathrm{d}V$, and the so-called 
boxcar function ${\cal{B}}$ is defined as
\begin{equation}
  {\cal{B}}(u, v, \delta) = 
     \left\{\begin{array}{ll}
       1 & |u-v|<\frac{\delta}{2}\\
       0 & \mathrm{otherwise}\\
     \end{array}\right.
  \label{eq:boxcar_function}
\end{equation}
The resolution parameter $\delta$, related to the resolution of images ($400\times400$ pixels), 
was set to $6v_\mathrm{orb}/400$. 

\begin{figure*}[t]
  \centering 
  \resizebox{\hsize}{!}{
    \includegraphics{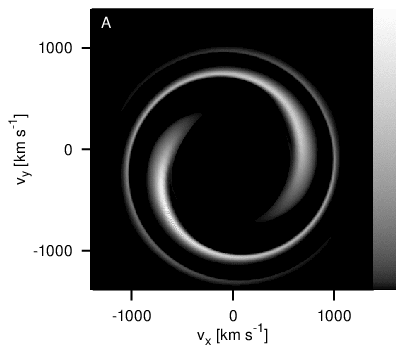}
    \includegraphics{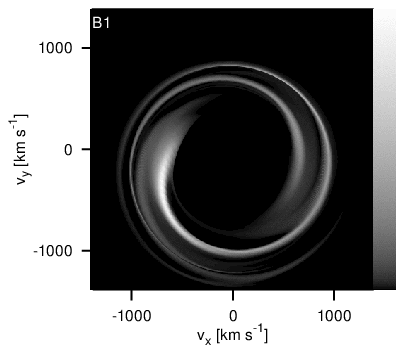}
  }
  \resizebox{\hsize}{!}{
    \includegraphics{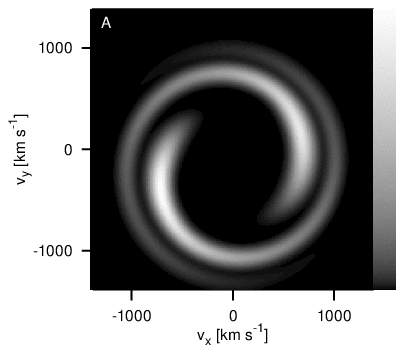}
    \includegraphics{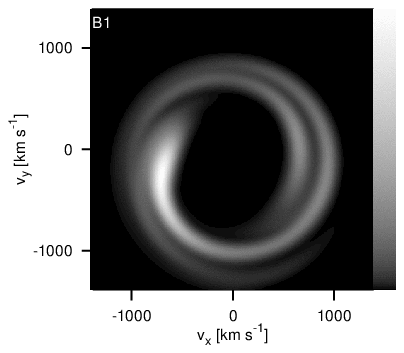}
  }
  \caption{Doppler tomograms of compressional power distributions shown in Fig.~\ref{miscmaps}. 
    Top row: raw data. Bottom row: the same data blurred by the convolution with a Gaussian:
    $exp\left[-(\frac{u_x-v_x}{10\delta_{v}})^2-(\frac{u_y-v_y}{10\delta_{v}})^2\right]$.}
  \label{dopplerpdv}
\end{figure*}

In the three-body approximation of Smak (2001), the inner disk is essentially Keplerian, and its 
Doppler tomogram cannot show any nonaxisymmetric structures in the high-velocity range. However, 
in a more realistic polytropic disk the spiral shocks excited at its outer edge propagate deeply 
into the high-velocity regions. As a result, our tomograms (Fig.~\ref{dopplerpdv}) show extended 
spirals instead of two crescent-shaped maxima reported by Smak (2001). Such spirals are not 
observed in real disks (Groot 2001). To account for the limited resolution of the observational 
data we blurred the tomograms to such a degree that the width of the brightest segments of the 
spiral became comparable to the width of the observed features 
($\sim400\div500\,\mathrm{km\,s^{-1}}$; see Groot 2001). However, the spiral pattern extending up 
to velocities of $\sim 1000\,\mathrm{km\,s^{-1}}$ was still clearly visible. Moreover, both 
location and relative intensity of the brightest segments of the spiral did not agree with those 
observed in U~Gem. 

Should this disagreement be regarded as an argument against the presence of spiral waves in CV 
disks? Certainly not. As we already indicated in the Introduction, the observed tomograms refer 
to the distribution of the emissivity in specific spectral lines rather than to the distribution 
of physical parameters directly obtainable from hydrodynamical simulations. The presently 
available hydrocodes are not sophisticated enough to predict the detailed spectrum of the disk, 
and the correspondence between those two sets of data is by no means clear. However, the models 
can yield data much more closely related to line emissivity than simple physical parameters or 
their combinations. 

To obtain such data, we assume the line emission to originate mainly due to the irradiation of the 
surface layer of the disk by the central white dwarf and/or boundary layer ({\it cf.} Robinson et 
al. 1993, Smak 1991). Since our models are not detailed enough to resolve the surface layer, we 
assume that the lower boundary of the layer coincides with a constant density surface 
$S_\mathrm{l}$ at which $\rho\equiv\rho_\mathrm{l}=10^{-8}$. Typical distance between 
$S_\mathrm{l}$ and the midplane of the disk, $h_\mathrm{l}$ was such, that $h_\mathrm{l}/r=0.08$ 
and $h_\mathrm{l}/r=0.2$ in the inner disk and in the outer ring, respectively. Further, we 
assume that the line flux from each element of the layer is proportional to the mass contained 
within that element, $\rho \mathrm{d}V$, multiplied by the irradiating flux. For simplicity, we 
also assume that all irradiating photons are emitted from a point source located at the centre of 
the white dwarf, so that the irradiating flux is given by ${\cal L}/r^2$, where $r$ is the radial 
coordinate of the volume element $\mathrm{d}V$, and 
${\cal L}={\cal L}_\mathrm{wd}+{\cal L}_\mathrm{bl}$ is the combined luminosity of the white 
dwarf and the boundary layer. The distribution of brightness on the $(v_x,v_y)$ plane is now 
given by the integral
\begin{equation}
  L_\mathrm{irr}(v_x,v_y) = \int_V\rho\frac{\cal L}{r^2} {\cal{S}}(r,\theta,\phi) 
  {\cal{B}}(u_x, v_x, \delta_{v_x}) {\cal{B}}(u_y, v_y, \delta_{v_y}) \mathrm{d}V.
  \label{eq:irrad_tomogram}
\end{equation}
The function ${\cal{S}}(r,\theta,\phi)$ describes the shadow cast by $S_\mathrm{l}$, and it is 
given by 
\begin{equation}
  {\cal{S}}(r,\theta,\phi) = 1 - {\cal{H}} \left(\int_0^r {\cal{H}}
  (\rho(r',\theta,\phi)-\rho_\mathrm{l}) dr'\right), 
  \label{eq:shadow_function}
\end{equation}
where 
\begin{equation}
  {\cal{H}}(x) =
     \left\{\begin{array}{ll}
              1 & x>0\\
              0 & \mathrm{otherwise}\\
            \end{array}
     \right.
  \label{eq:heaviside_function}
\end{equation}
is the Heaviside step function. The third velocity component, $v_z$, was neglected in 
(\ref{eq:irrad_tomogram}) because nearly everywhere in the disk its value was smaller than 
$\sim5\%$ of the local azimuthal velocity, $v_{\phi}$. Formally, the integral in 
Eq.~\ref{eq:irrad_tomogram} subtends the whole space above $S_\mathrm{l}$. Practically, due to 
steeply falling density, only volume elements closest to $S_\mathrm{l}$ contribute to it 
significantly. The boundary layer above $S_\mathrm{l}$ has a mass of $\simeq0.15M_\mathrm{disk}$, 
but only about 25\% of its volume is directly illuminated.

\begin{figure*}[t]
  \centering 
  \resizebox{\hsize}{!}{
    \includegraphics{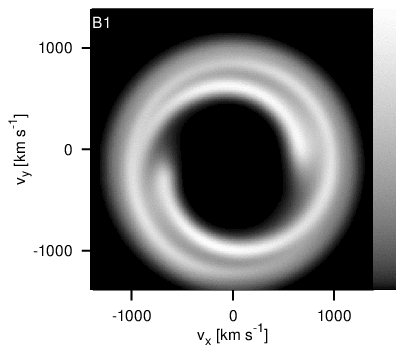}
    \includegraphics{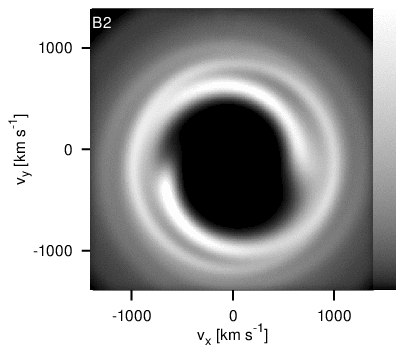}
  }
  \caption{Tomograms obtained from models~B1 and~B2 within the irradiation approach (blurred in 
    the same way as described in Fig.~\ref{dopplerpdv}).}
  \label{dopplerirrad}
\end{figure*}

\begin{figure*}[t]
  \centering 
  \resizebox{\hsize}{!}{
    \includegraphics{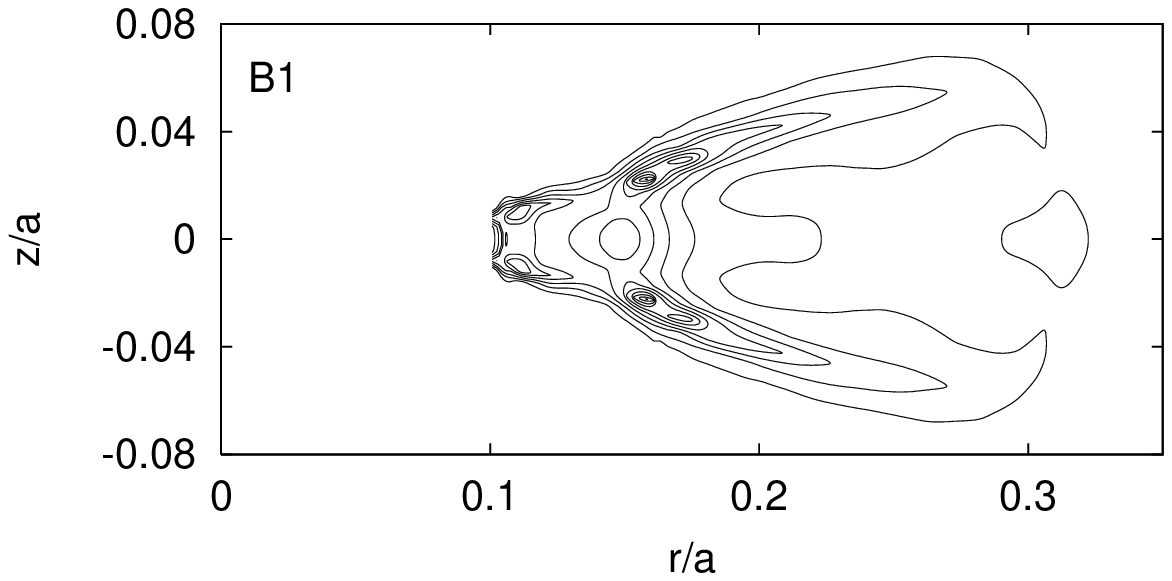}
    \includegraphics{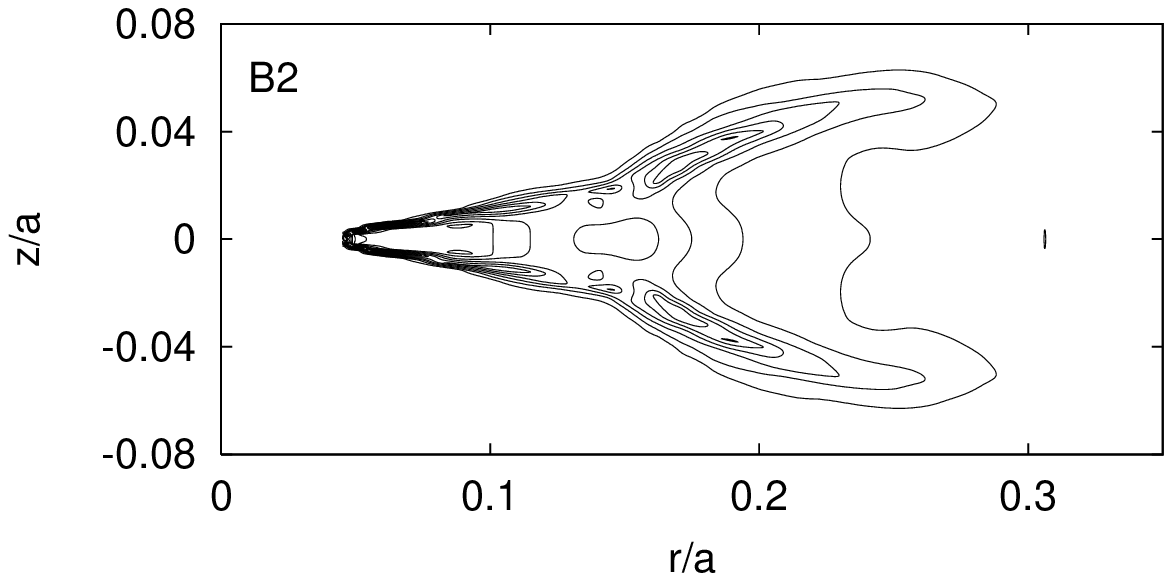}
  }
  \caption{Distribution of the compressional power per unit mass averaged over the azimuthal angle,
    $-\int_0^{2\pi}min\left(\frac{p\nabla\vec{v}}{\rho},0\right) \mathrm{d}\phi$, in models~B1 
    and~B2.}
  \label{pdvravg}
\end{figure*}

Location and relative intensity of the brightest areas on tomograms  resulting from the 
irradiation approach (Fig.~\ref{dopplerirrad}) agree rather well with those observed by 
Groot (2001) at an advanced outburst phase of U~Gem (his Fig.~2, Episode~2). The major 
discrepancy is the bright ring visible in our tomograms at 
$\sqrt{v_x^2+v_y^2} \simeq 1000\,\mathrm{km\,s^{-1}}$. In both models the ring is too bright 
compared to the observational data, but relative to the maximum intensity obtained in the model 
it is weaker in B2 where the disk extends down to $r=0.045$. We conclude that in B1 the ring is 
enhanced by a spurious contribution from the inner edge of the disk at $r=0.1$. The ring in B2 
would be still weaker if absorption of the irradiating flux by the gas in the surface layer was 
taken into account in equation (\ref{eq:irrad_tomogram}). Unfortunately, the present models are 
too crude for such an operation to be reliable. It is clear, however, that the effect of 
absorption should be particularly strong for $r\lesssim0.1$, where $h/r$ is nearly constant (see 
Fig.~\ref{pdvravg}), and the irradiating quanta propagate nearly parallel to $S_\mathrm{l}$. 
Further reduction in ring intensity could probably be achieved if the inner boundary of the grid 
was moved even closer to the white dwarf, as the shadow cast by $S_\mathrm{l}$ at $r<0.045$ might 
partly screen the region at $r\sim0.1$ where the ring originates. On the other hand, in some 
phases of the activity cycle the intensity of the brightest areas in our irradiation tomograms is 
underestimated relative to the ring. This is because substantial $p\mathrm{d}V$ work is done by 
tidal forces directly on the gas in the surface layer of the outer disk, where the low-velocity 
emission originates. 

In fact, the heating rate per unit mass, $p\mathrm{d}V/\rho$, reaches a clear maximum just below 
the surface of the outer disk (see Fig.~\ref{pdvravg}). For the case of U~Gem the height of this 
maximum is $\simeq10^{11}\,\mathrm{erg\,g^{-1}\,s^{-1}}$. With a typical outburst accretion rate 
of $\dot{M}=3\cdot10^{18}\,\mathrm{g\,s^{-1}}$, and a white dwarf radius 
$R_\mathrm{wd}=4\cdot10^{8}\,\mathrm{cm}$ we get 
$L_\mathrm{bl}\simeq{\cal L}=\frac{1}{2}G M_1 \dot{M}/R_\mathrm{wd}\simeq6\cdot10^{35}\,\mathrm{erg\,s^{-1}}$.
The illuminated mass in the region of maximum $p\mathrm{d}V/\rho$ between $r\sim0.15a$ and 
$r\sim0.25a$ approaches $\simeq0.025M_\mathrm{disk}$, and it is distributed within a solid angle 
of $\sim0.3\pi$. Assuming that the whole incident flux is absorbed there we obtain a radiative 
energy input of $\sim2\cdot10^{12}\,\mathrm{erg\,g^{-1}\,s^{-1}}$, and we see that during the 
outburst the contribution to the line flux from $p\mathrm{d}V$ heating is rather small. However, 
$p\mathrm{d}V$ dominates just before the outburst, when accretion rate is factor of $\sim100$ 
lower.

The subsurface maximum of $p\mathrm{d}V/\rho$ in Fig.~\ref{pdvravg} originates mainly due to tidal 
forcing in the direction perpendicular to the orbit. It also contains a contribution from the 
ambient gas falling onto the disk; however the ``rainfall'' heating is much less efficient than 
the tidal one. We checked this by re-running simulation B1 with $\rho_\mathrm{min}$ reduced to 
$3\cdot10^{-14}$: at $t\simeq8.0\,P_\mathrm{orb}$ virtually no changes were seen in the 
distribution of $p\mathrm{d}V/\rho$. 

\section{Discussion}
\label{disc}

As discussed in Section~\ref{results}, we find that spiral waves efficiently propagate from 
excitation regions at the outer edge of the disk toward the white dwarf, reaching to at least 
$\sim0.05a$. This conclusion concerns both two-dimensional and three-dimensional models; it is 
however limited to hot polytropic disks presented in this paper. The main spiral features seen in 
the density distribution (two-dimensional case) and in the surface density distribution 
(three-dimensional case) are hard to distinguish. On the other hand, clear differences are 
visible in the distributions of the tidal heating rate, $p\mathrm{d}V$: in 3D the two main spiral 
arms are less tightly wound than in 2D, and a weaker third arm is excited. We suggest that the 
third arm may originate from tidal forcing in the direction perpendicular to the orbital plane. 
The effects of this forcing seem to be responsible for the origin of the clear maximum of tidal 
heating rate per unit mass, $p\mathrm{d}V/\rho$, which is located away from the midplane in 
subsurface layers of the outer disk. 

Doppler tomograms of tidal heating rate derived from both 2D and 3D models correlate rather 
poorly with observed tomograms of U~Gem (Groot 2001). A better agreement (but still not entirely 
satisfactory) is obtained for tomograms of the irradiation flux from the white dwarf through the 
surface layer of the disk. The brightest areas of such tomograms coincide with arches observed in 
U~Gem at an advanced stage of the outburst. The irradiation tomograms can be derived from 3D 
models only, which indicates that fully three-dimensional modelling is needed for a reliable 
interpretation of the observational data on DN disks. 

According to our results the arches originate in the outer part of the disk, fairly high above 
the midplane ($h>0.1r$). For this to happen, the outer disk would have to be substantially 
bulged. The bulging phenomenon may be explained within the following (not entirely new) scenario: 
prior to the outburst the gas transferred from the secondary mainly collects in a ring at the 
circularization radius, and only partly accretes through the disk onto the white dwarf. The ring 
expands as the gas flows in, but it remains cool until heating from tidal forcing in the orbital 
plane grows so strong that it cannot be balanced by radiative cooling. The ring begins to expand 
even more rapidly, and within it the gas located away from the midplane begins to receive 
additional internal energy from tidal forcing in the direction perpendicular to the orbital 
plane. Eventually, strong spiral shocks develop, and a dynamical instability of the kind 
described by R\'o\.zyczka \& Spruit (1993) sets in. 

Obviously, such scenario cannot be the whole story, as it is not linked to the thermal instability 
believed to be at least partly responsible for eruptive phenomena in DN and other classes of 
cataclysmic variables. Nevertheless, it seems to indicate a promising direction of further 
research. 

\medskip
\noindent
Acknowledgments. We are greatly indebted to Professor J. Smak for enlightening discussions and 
critical reading of the manuscript.
This work was supported by the Committee for Scientific Research through the grant 2P03D 01419.

\medskip
\noindent
References

\noindent
Boffin, H.~M.~J. 2001, ``Astrotomography. Indirect Imaging Methods in Astronomy'', Eds. H.~M.~J. Boffin, D. Steeghs, J. Cuypers, Lecture Notes in Physics 573, p.69\\
Clarke, D.~A. 1996, Astrophys. J., 457, 291\\
Clarke, D.~A. and Norman, M.~L., 1994, LCA preprint 7 (Urbana-Champaign: Univ. Ill.)\\
Groot, P.~J. 2001, Astrophys. J., 551, L89\\
Kley, W. 1998, Astron. Astrophys., 338, L37\\
Marsh, T.~R. 2001, ``Astrotomography. Indirect Imaging Methods in Astronomy'',  Eds. H.~M.~J. Boffin, D. Steeghs, J. Cuypers, Lecture Notes in Physics 573, p.1\\
Robinson, E.~L., Marsh, T.~R. and Smak, J. 1993, ``Accretion Disks in Compact Stellar Systems'', Ed. J.~C. Wheeler (Singapore: World Scientific), p. 75\\
R\'o\.zyczka, M., and Spruit, H. 1993, Astrophys. J., 417, 677\\
Smak, J. 1991, ``Structure and Emission Properties of Accretion Disks'', Eds. C. Bertout, S. Collin-Souffrin, J.~P. Lasota, J. Tran Thanh Van (Paris: Editions Fronti\'eres), p. 247\\
Smak, J., 2001 Acta Astron., 51, 295\\
Steeghs, D. and Stehle, R. 1999, MNRAS, 307, 99\\

\end{document}